\documentclass[aps,prl,twocolumn,superscriptaddress]{revtex4}
\usepackage[dvips]{graphicx}
\newcommand{\ket}[1]{|#1\rangle}             
\newcommand{\bra}[1]{\langle#1|}             
\newcommand{\op}[1]{\widehat{#1}}            
\begin{document}
\title{Spin-orbit hybrid entanglement of photons and quantum contextuality}
\author{Ebrahim Karimi}
\email{karimi@na.infn.it}
\affiliation{Dipartimento di Scienze Fisiche, Universit\`{a} di
Napoli ``Federico II'', Compl.\ Univ.\ di Monte S. Angelo, 80126
Napoli, Italy}
\author{Jonathan Leach}
\affiliation{Department of Physics and Astronomy, University of
Glasgow, Glasgow, Scotland, UK}
\author{Sergei Slussarenko}
\affiliation{Dipartimento di Scienze Fisiche, Universit\`{a} di
Napoli ``Federico II'', Compl.\ Univ.\ di Monte S. Angelo, 80126
Napoli, Italy}
\author{Bruno Piccirillo}
\affiliation{Dipartimento di Scienze Fisiche, Universit\`{a} di
Napoli ``Federico II'', Compl.\ Univ.\ di Monte S. Angelo, 80126
Napoli, Italy}
\affiliation{Consorzio Nazionale Interuniversitario per le Scienze
Fisiche della Materia, Napoli}
\author{Lorenzo Marrucci}
\affiliation{Dipartimento di Scienze Fisiche, Universit\`{a} di
Napoli ``Federico II'', Compl.\ Univ.\ di Monte S. Angelo, 80126
Napoli, Italy}
\affiliation{CNR-SPIN, Compl.\ Univ.\ di Monte S. Angelo, 80126
Napoli, Italy}
\author{Lixiang Chen}
\affiliation{State Key Laboratory of Optoelectronic Materials and
Technologies, Sun Yat-sen University, Guangzhou 510275, China}
\author{Weilong She}
\affiliation{State Key Laboratory of Optoelectronic Materials and
Technologies, Sun Yat-sen University, Guangzhou 510275, China}
\author{Sonja Franke-Arnold}
\affiliation{Department of Physics and Astronomy, University of
Glasgow, Glasgow, Scotland, UK}
\author{Miles J. Padgett}
\affiliation{Department of Physics and Astronomy, University of
Glasgow, Glasgow, Scotland, UK}
\author{Enrico Santamato}
\affiliation{Dipartimento di Scienze Fisiche, Universit\`{a} di
Napoli ``Federico II'', Compl.\ Univ.\ di Monte S. Angelo, 80126
Napoli, Italy}
\affiliation{Consorzio Nazionale Interuniversitario per le Scienze
Fisiche della Materia, Napoli}

\begin{abstract}
We demonstrate electromagnetic quantum states of single photons and
of correlated photon pairs exhibiting ``hybrid'' entanglement
between spin and orbital angular momentum. These states are obtained
from entangled photon pairs emitted by spontaneous parametric down
conversion, by employing a $q$-plate for coupling the spin and
orbital degrees of freedom of a photon. Entanglement and contextual
quantum behavior (that is also non-local, in the case of photon
pairs) is demonstrated by the reported violation of the
Clauser-Horne-Shimony-Holt inequality. In addition a
classical analog of the hybrid spin-orbit photonic entanglement is
reported and discussed.
\end{abstract}
\maketitle

\section{Introduction}
Entangled states are at the heart of most quantum paradoxes and
provide the main tool for quantum information processing, including
applications such as teleportation, cryptography, superdense coding,
etc. Entangled quantum states are also the basis of Bell's
inequality violations, which ruled out classical hidden-variable
theories in favor of quantum mechanics~\cite{bell66}. Bell's
inequalities were originally derived for two particles, as a
consequence of locality and realism. In almost all experimental
demonstrations of these inequalities to date, the same degree of
freedom of two particles has been used, e.g. the spin of a photon.
Very recently, however, the case of so-called ``hybrid
entanglement'', occurring when the involved degrees of freedom of
the two particles are not the same, has attracted a certain
interest, and the first experimental demonstrations with spin and
spatial-mode degrees of freedom have been reported
\cite{ma09,neves09}. Using different degrees of freedom also opens
up another opportunity, i.e. that of realizing entanglement between
different degrees of freedom of a single particle. In this case, no
role is played by non-locality, but Bell-type inequalities can still
be formulated by assuming realism and the so-called
``non-contextuality'' of the two involved commuting observables,
i.e. the assumption that the result of a particular measurement of
one observable is determined independently of any simultaneous
measurement of the other one~\cite{roy93,mermin93,liu09}.
Non-contextual hidden variable models have been excluded by recent
experiments where the violation of suitable inequalities was
observed using neutrons~\cite{hasegawa03}, ions \cite{kirchmair09},
and single photons prepared in entangled spin-path
states~\cite{gadway09}.  Finally, single-particle entanglement, in
the case of bosons such as photons, has a classical analog that is
obtained by replacing single-photon states with multi-photon
coherent states realized within the same field mode
\cite{spreeuw98}. Such a classical analog helps visualizing the
nature of the single-particle entanglement.

A particularly convenient framework in which to explore these
concepts is provided by photons carrying both spin angular momentum
(SAM) and orbital angular momentum (OAM). While the former is the
most widely employed internal degree of freedom of photons for
quantum manipulations, the latter is becoming an interesting
additional resource for quantum applications (see, e.g.,
\cite{molina07,franke08,barreiro08,nagali09a}). In this work, we
study three conceptually related experimental situations. Firstly,
heralded single photons are prepared in a state where SAM and OAM
are entangled (as proposed in \cite{chen10}), and are then used for
testing the contextuality of different degrees of freedom of the
same particle. Secondly, correlated photon pairs, where the SAM of
one photon is entangled with the OAM of the other, i.e. photon pairs
exhibiting SAM-OAM hybrid entanglement, are generated and used for
testing the contextuality and non-locality of these degrees of
freedom when they are spatially separated. Finally, optical coherent
states involving many photons, are used to demonstrate a classical analog of
SAM-OAM hybrid entanglement.
\begin{figure*}[!htbp]
\centering
\includegraphics[width=11cm]{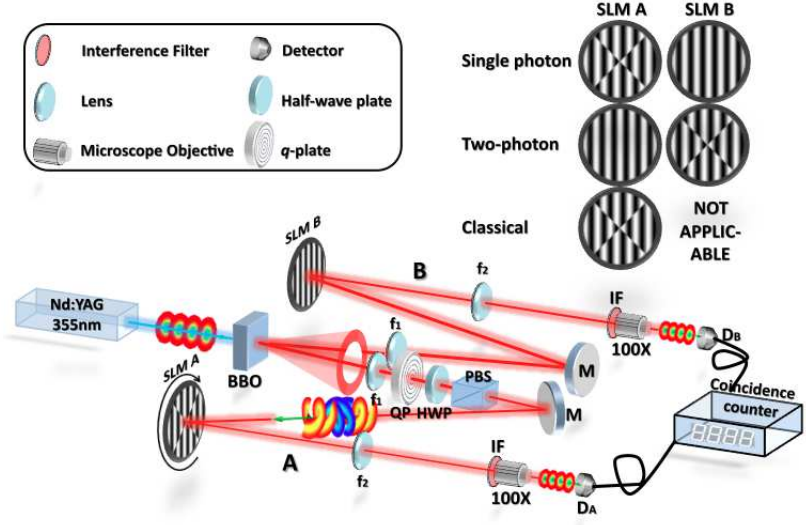}
    \caption{ \label{setup}
Setup used for the two quantum-regime experiments. A Nd:YAG laser
with average power of 150 mW at 355 nm pumps a nonlinear crystal of
$\beta$-barium borate (BBO) cut for degenerate type-I
non-collinear phase matching which emits OAM-entangled $H$-polarized
photon pairs at 710 nm (see Ref.\ \cite{leach09} for details). The
photons of each pair are split in arms \textbf{A} and \textbf{B},
respectively. Legend of the main components (see also graphic symbol
legend in the upper-left inset): f$_1$, f$_2$ - lenses for beam
control; QP - $q$-plate; HWP - half-wave plate; PBS - polarizer; M -
mirror; SLM A and SLM B - spatial light modulators; IF -
interference filter for bandwidth definition; 100X - microscope
objectives for fiber coupling; D$_A$, D$_B$ - photon detectors. In
the classical-regime experiment, the optical line is the same as arm
\textbf{A}. Top-right inset: computer-generated hologram patterns
displayed on the two SLMs in the three experiments.}
\end{figure*}

\section{Experimental Setup}
The experimental layout we used in the quantum regime (the first two
experiments) is presented in Fig.\ \ref{setup}. Our down-conversion
source generates photon pairs that are entangled in the OAM degree
of freedom \cite{jack09,leach09}, each photon being horizontally
polarized, as described by
\begin{equation}\label{eq:spdc_state}
    \ket{\psi}= \sum_{m=-\infty}^{\infty}c_{|m|}\ket{m}_{o}^{A}\ket{-m}_{o}^{B}
    \ket{H}^A_\pi\ket{H}^B_\pi.
\end{equation}
%
%
Here $A$ and $B$ denote the signal and idler photons traveling along
the two corresponding arms of the setup shown in Fig.\ \ref{setup}
and $\pi$, ${o}$ denote SAM and OAM degrees of freedom,
respectively. The integer $m$ is the photon OAM in units of $\hbar$
and $H$ denotes horizontal linear polarization.

\subsection{Single-photon experiment}
\begin{figure}[ht]
\centering
\includegraphics[width=8.5cm]{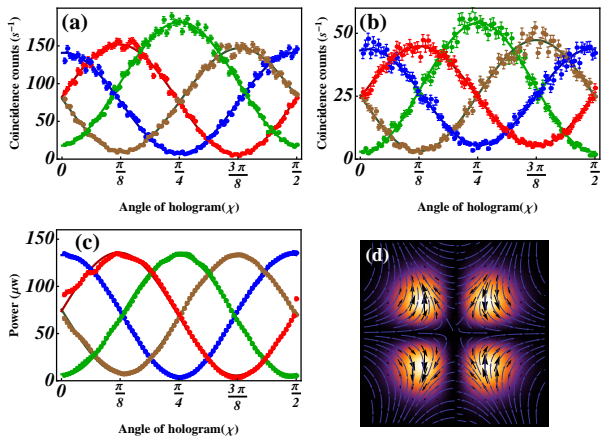}
\caption{\label{coincidence_count}
    The experimental coincidence counts as a function of orientation of the sector hologram
    for different values of polarization direction, for heralded single photons (a),
    photon pairs (b) and coherent-states (c): blue dots - $\theta=0$, red dots - $\theta=\pi/4$,
    green dots - $\theta=2\pi/4$, gray dots - $\theta=3\pi/4$. The solid lines are the best theoretical fit over
    the experimental data. The fringe contrast is about 90\%, which is much larger
    than 70.7\%, as required for Bell's inequality verification.
    (d) Simulated intensity and polarization distribution patterns of the optical field for
    the beam emerging from the $q$-plate in the case of horizontal polarization input beam.}
\end{figure}

\begin{figure}[ht]
\centering
\includegraphics[width=6.cm]{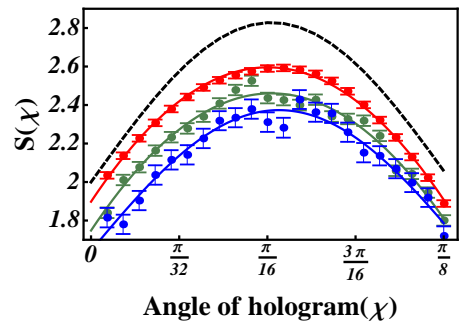}
    \caption{\label{s_sam_oam_value}
    The CHSH $S$ value in a region where it is larger than the classical limit
    $2$. The choice of the variables appearing in Eq.\ (\ref{eq:s}) is the following: $\theta=0$,
    $\theta'=\pi/4$, $\chi$ is the plot abscissa, $\chi'=\chi+\pi/8$.
    The green, blue, and red dots correspond to the experimental data in the case of single-photon (a), photon-pairs (b),
    and classical-wave (c) SAM-OAM experiments, respectively. The dashed line is the quantum mechanical
    ideal prediction.  In the two cases (a) and (b), at $\chi=\pi/16$, the CHSH inequality is violated respectively by
    17 and 10  standard deviations.  The classical case (c) is plotted for comparison}
\end{figure}

. 
In this case we use photon $B$
to herald a single photon $A$ which we convert into an OAM-SAM
maximally entangled state. Starting from state $\ket{\psi}$ given in
Eq.~(\ref{eq:spdc_state}), we post-select photon pairs having $m=0$
, i.e. in state
$\ket{\psi}=c_0\ket{0}_{o}^{A}\ket{0}_{o}^{B}\ket{H}^A_\pi\ket{H}^B_\pi$,
by coupling photon $B$ into a single-mode optical fiber. Photon $A$
is thus also projected into $m=0$. Spatial light modulator SLM B in
this case is patterned as a uniform grating, deflecting the beam but
not affecting its transverse spatial mode (see upper-right inset of
Fig.\ \ref{setup}). Photon $A$ is sent first through a $q$-plate
\cite{marrucci06,marrucci06b} so as to generate the maximally
entangled SAM-OAM state~\cite{nagali09}
\begin{equation}\label{eq:qp}
    \ket{\Phi^{+}}^A=\frac{1}{\sqrt{2}}
    \left(\ket{R}^A_\pi\ket{+2}^A_{o}+\ket{L}^A_\pi\ket{-2}^A_{o}\right)
\end{equation}
where $L$, $R$ denote left-circular and right-circular polarization
states, respectively. The polarization state of photon $A$ emerging
from the $q$-plate is then measured by a half-wave plate (HWP)
oriented at a variable angle $\theta/2$ and a fixed linear
polarizer, restoring the horizontal polarization. This HWP-polarizer
combination filters incoming photons having linear polarization at
angle $\theta$ with respect to the horizontal direction. In the
circular polarization basis, the state of the filtered photons is
written as
$\ket{\theta}_\pi=\frac{1}{\sqrt{2}}\left(e^{i\theta}\ket{L}_\pi +
e^{-i\theta}\ket{R}_\pi\right)$. The SAM measurement does not affect
the OAM degree of freedom. Non-contextuality can be assumed between
the $z$-component of photon SAM and OAM, because, in the paraxial
approximation, the SAM operator $\op{S}_z$ commutes with the OAM
operator $\op{L}_z$. After SAM filtering, the photon's OAM is also
measured by a suitable computer-generated hologram, displayed on SLM
A, followed by coupling into a single-mode fiber. The hologram
pattern is defined by the four-sector alternated $\pi$-shift phase
structure shown in the upper-right inset of Fig.\ \ref{setup}, with
the four sectors rotated at a variable angle $\chi$ (the grating
fringes are not rotated). On diffraction, this hologram transforms
the photons arriving in the OAM superposition state
$\ket{\chi}_{o}=\frac{1}{\sqrt{2}}\left(e^{2i\chi}\ket{+2}_{o} +
e^{-2i\chi}\ket{-2}_{o}\right)$ back into the $m=0$ state, which is
then filtered by coupling in fiber. The OAM superposition state
$\ket{\chi}_{o}$ is the spatial mode analog of the linear
polarization, and we may refer to its angle $\chi$ as to its
``orientation'' \cite{footnote1}. The overall effect of our
apparatus is therefore to perform a joint measurement of the
polarization and spatial mode orientations of $A$ photons at angles
$\theta$ and $\chi$, respectively. When  photon $A$ is in the
entangled Bell state described by Eq.~(\ref{eq:qp}), we expect that
the final probability to detect it (in coincidence with the $B$
trigger photon) is given by
\begin{equation}\label{eq:prob}
    P({\theta,\chi})=|^A\bra{\Phi^{+}}\cdot\ket{\theta}^A_\pi\ket{\chi}^A_{o}|^2
    \propto \cos^2{(\theta-2\chi)}.
\end{equation}
To test entanglement we adopt the Clauser-Horne-Shimony-Holt (CHSH)
inequality, given by \cite{CHSH}
\begin{eqnarray}\label{eq:s}
    S=|E(\theta,\chi)-E(\theta,\chi')+E(\theta',\chi)+E(\theta',\chi')|\le 2,
\end{eqnarray}
where $E(\theta,\chi)$ is calculated from the $A$-$B$ photon
coincidence counts $C(\theta,\chi)$ according to
\begin{widetext}
\begin{eqnarray}\label{eq:c_n}
    E(\theta,\chi)=\frac{C(\theta,\chi)+C(\theta+\frac{\pi}{2},\chi+\frac{\pi}{4})-C(\theta+\frac{\pi}{2},\chi)-C(\theta,\chi+\frac{\pi}{4})}
    {C(\theta,\chi)+C(\theta+\frac{\pi}{2},\chi+\frac{\pi}{4})+C(\theta+\frac{\pi}{2},\chi)+C(\theta,\chi+\frac{\pi}{4})}.
\end{eqnarray}
\end{widetext}
Whilst the CHSH inequality is commonly applied to non-local
measurements on two spatially separated entangled photons, testing
for hidden variable theories, here we apply it to single-photon
entanglement to test for contextuality. In
Fig.~\ref{coincidence_count}a the coincidence counts are shown as a
function of spatial mode orientation $\chi$ for different values of
polarization angles $\theta$.
The occurrence of high-visibility fringes indicates
(single-particle) entanglement in the SAM-OAM spaces. The CHSH $S$
value calculated from this data is shown in Fig.
\ref{s_sam_oam_value} (green dots). A violation of the CHSH
inequality is clearly obtained, in good agreement with quantum
theory predictions, confirming the entanglement and providing a
demonstration of quantum SAM-OAM contextuality for single photons.

\subsection{Two-photon experiment}. 
In this case, we generate and
verify entanglement between the SAM of one photon and the OAM of the
other, i.e. we demonstrate non-local hybrid entanglement in these
two degrees of freedom. To this purpose, the four-sector and uniform
holograms of arms \textbf{A} and \textbf{B} were swapped, as
displayed in the top-right inset of Fig.~\ref{setup}. The $q$-plate
in arm \textbf{A} and the sector hologram in arm \textbf{B} of the
apparatus, together with subsequent coupling into the single-mode
fiber before detection, act so as to post-select the photons with
$m=\pm2$ in Eq.~(\ref{eq:spdc_state}), i.e. the post-selected
initial two-photon state is
$\ket{\psi}=\frac{1}{\sqrt{2}}c_2\left(\ket{2}_{o}^{A}\ket{-2}_{o}^B +
\ket{-2}_{o}^A\ket{2}_{o}^B\right)\ket{H}^A_\pi\ket{H}^B_\pi$. The
 photon $A$ passes through the $q$-plate, acting in this case as a
OAM-to-SAM transferrer \cite{nagali09}, so that the OAM eigenstates
$m=\pm2$ are mapped into $L$ and $R$ polarized photons with $m=0$,
respectively. After this process, the photon pair is projected into
the nonlocal state
\begin{equation}\label{eq:phi_nonlocal}
   \ket{\phi}_{nl}=\frac{1}{\sqrt{2}}\left(\ket{L}_\pi^A\ket{+2}_{o}^B+
\ket{R}_\pi^A\ket{-2}_{o}^B\right)\ket{0}^A_{o}\ket{H}^B_\pi
\end{equation}
where the SAM of one photon is maximally entangled with the OAM of
the other. Next, the polarization of the $A$ photon is measured by
the HWP rotated at angle $\theta/2$ followed by the polarizer, and
the spatial mode of the $B$ photon by the sector hologram rotated at
angle $\chi$ followed by coupling in fiber. Well-defined coincidence
fringes with visibility up to 90\% are obtained, as shown in
Fig.~\ref{coincidence_count}b. Repeating the measurements for
different angles $\theta$ and $\chi$, the quantity $S$ was evaluated
from Eqs.~(\ref{eq:s}) and (\ref{eq:c_n}) and the violation of the
CHSH inequality was verified, as shown in Fig.~\ref{s_sam_oam_value}
(blue dots). This violation provides a demonstration of SAM-OAM hybrid
entanglement and non-locality, for separated photon pairs.

\subsection{Classical light experiment}. 
In our final experiment, we
move to a classical regime of non-separable optical modes occupied
by many photons, corresponding to coherent quantum states. A 100 mW
frequency-doubled linearly-polarized continuous wave Nd:YVO$_4$
laser beam is sent in an optical line equal to arm \textbf{A} of our
quantum apparatus, so as to obtain, after the $q$-plate, a coherent
state in the SAM-OAM non-separable mode $\ket{\Phi^{+}}$ given by
Eq.~(\ref{eq:qp}) \cite{footnote2}. The calculated structure of this
mode is shown in Fig.~\ref{coincidence_count}d, for a given input
polarization. The mode non-separability is evident, as the
polarization is spatially non-uniform \cite{borges10}. The beam
polarization is then filtered by the combination of the HWP at
angle $\theta$ and polarizer and its spatial mode by the sector
hologram rotated at angle $\chi$, as in the single-photon experiment
(a). In this case, no trigger is used and the count rates
$C(\theta,\chi)$ in Eq.~(\ref{eq:c_n}) are replaced by average power
measurements, corresponding to photon fluxes. When the angles
$\theta$ and $\chi$ are changed, high contrast sinusoidal fringes
proportional to $\cos^2(\theta-2\chi)$ were observed in the overall
transmitted power fraction, as shown in Fig.\
\ref{coincidence_count}c.  As shown in
Fig.~\ref{s_sam_oam_value} (red dots) we note that the classical
experiment  mimics the results of the single photon experiment.
However, the experiment can of course also be
interpreted without assuming the existence of photons. In this case,
SAM and OAM measurements can be understood just as wave filtering
procedures, and no conclusion can be drawn about discrepancies
between classical-realistic and quantum behaviour. Nevertheless,
providing a classical analog of single-particle entanglement is
interesting in itself and may offer the basis for some entirely
classical implementations of quantum computational tasks
\cite{spreeuw01}.

\section{Conclusions}
In conclusion, we have demonstrated hybrid entanglement between the
spin and the orbital angular momentum of light in two different
regimes: single photons and entangled photon pairs.  We have
reported an additional classical experiment which mimics the quantum
result and although the experimental results appear very similar in
the three cases, they provide different and complementary insight
into the contextual quantum nature of light. 

\section{Acknowledgment}
The project PHORBITECH
acknowledges the financial support of the Future and Emerging
Technologies (FET) programme within the Seventh Framework Programme
for Research of the European Commission, under FET-Open grant number
255914.


\begin{thebibliography}{22}
\bibitem{footnote2}
We note that, when the classical states are described as coherent
quantum states, the state of the non-separable mode
$\ket{\alpha_{\Phi^+}}$ is found to be decomposable in the direct
product of the coherent states of the two SAM-OAM eigenmodes
$\ket{\frac{1}{\sqrt{2}}\alpha_{(R,2)}}$ and
$\ket{\frac{1}{\sqrt{2}}\alpha_{(L,-2)}}$, so that no SAM-OAM photon
entanglement is actually present in the beam
\protect\cite{borges10}.
\end{thebibliography}

\end{document}